# Shear recovery and temperature stability of Ca²⁺ and Ag⁺ glycolipid fibrillar metallogels with unusual β-sheet-like domains

Alexandre Poirier,[a] Patrick Le Griel,[a] Thomas Bizien,[b] Thomas Zinn,[c] Petra Pernot,[c] Niki Baccile[a,*]

[a] Sorbonne Université, Centre National de la Recherche Scientifique, Laboratoire de Chimie de la Matière Condensée de Paris, LCMCP, F-75005 Paris, France
[b] Synchrotron Soleil, L'Orme des Merisiers, Saint-Aubin, BP48, 91192 Gif-sur-Yvette Cedex, France
[c] ESRF – The European Synchrotron, CS40220, 38043 Grenoble, France

* Corresponding author:
Dr. Niki Baccile
E-mail address: niki.baccile@sorbonne-universite.fr
Phone: +33 1 44 27 56 77

## Abstract

Low-molecular weight gelators (LMWG) are small molecules (Mw < ~1 kDa), which form self-assembled fibrillar networks (SAFiN) hydrogels in water. The great majority of SAFiN gels is described by an entangled network of self-assembled fibers, in analogy to a polymer in a good solvent. Here, fibrillation of a biobased glycolipid bolaamphiphile is triggered by Ca²⁺ or Ag⁺ ions, added to its diluted micellar phase. The resulting SAFiN, which forms hydrogel above 0.5 wt%, has a "nano-fishnet" structure, characterized by a fibrous network of both entangled fibers and β-sheets-like rafts, generally observed for silk fibroin, actin hydrogels or mineral imogolite nanotubes, but generally not known for SAFiN. This work focuses on the strength of the SAFIN gels, their fast recovery after applying a mechanical stimulus (strain) and their unusual resistance to temperature, studied by coupling rheology to small angle X-ray scattering (rheo-SAXS) using synchrotron radiation. The Ca²⁺-based hydrogel keeps its properties up to 55°C, while the Ag⁺-based gel shows a constant elastic modulus up to 70°C, without appearance of any gel-to-sol transition temperature. Furthermore, the glycolipid is obtained by fermentation from natural resources (glucose, rapeseed oil), thus showing that naturally-engineered compounds can have unprecedented properties, when compared to the wide range of chemically derived amphiphiles.

**Introduction**

Hydrogels are broadly intended as a highly branched network able to retain a large water content, of utmost importance for medicine and more generally for hygiene and medical product development, among many other applications.[1,2] Water thickening is generally produced by an entangled fibrous network of polymers, chemically or physically cross-linked,[3] giving rise to widespread consumer products.[4] To improve biocompatibility, biopolymer-based hydrogels became a source of much interest, largely studied up to now.[5] However, gelling by polymers is often an irreversible process, reason for which low molecular weight gelators (LMWG) are developed, for they combine a fascinating self-assembly behaviour, driven by non-covalent interactions, and a supramolecular assembly process, which can be tuned[6,7] more or less reversibly to form a gel.[8–10] However, may they be polymeric or self-assembled fibrillar networks (SAFiN), physical gels have in common the same entangled structure, identified as bundled flexible unidimensional fibrils under dilute or semi-dilute conditions.[11]

In this work, we report the formation of SAFiN hydrogels having a unique "nano-fishnet" (Figure 1d) β-sheet structure, similarly to silk fibroin[12–15] or actin.[16–18] They are entirely composed of a bolaform glycolipid with an oleic acid, C18:1, backbone. This compound, G-C18:1 (Figure 1a), displays a triple surfactant-lipid-gelator nature.[19] Below neutral pH and at concentrations below 5 wt%, G-C18:1 forms vesicles (lipid-like behaviour),[20,21] above neutral pH it forms micelles (surfactant behaviour),[20,21] while in the presence of micelles and metal ions, it forms wormlike micelles or fibers, according to the chemical nature of the ion.[22] The micelle-to-fiber phase transition is triggered at pH above neutrality by addition of specific cations ($Ca^{2+}$, $Ag^+$, $Mn^{2+}$).[22] Cation-induced fibrillation is not only unknown for this compound and other biosurfactants,[23–25] but it is actually not expected for surfactant solutions in general, which rather undergo micelle-to-cylinder,[26–29] -wormlike,[30] -vesicle[31] or -lamellar[29] transitions when they are mixed with mono- or multivalent cations. Increased viscosity of cation-surfactant solutions is not uncommon, although rare and attributed to wormlike structures.[32] On the contrary, metal-induced fibrillation is well-known for other classes of amphiphiles, such as low molecular weight gelators,[33,34] or peptide-based amphiphiles.[35,36]

Here, the cations do not only trigger fibrillation, but they also drive the side-by-side association of fibers into β-sheet-like rafts, which are characterized in a parallel work[37] by a combination of cryogenic transmission electron microscopy (cryo-TEM) and small-angle X-ray scattering (SAXS). Fiber rafts are themselves rarely observed for fibrous systems. Found in mineral imogolite (mineral aluminosilicates) nanotube systems,[38] they are in fact extremely rare for SAFiN[39,40] and, to the best of our knowledge, never reported within the context of SAFiN hydrogels (Table S 1).

Fibrillation and hydrogelation of glycolipids are not uncommon.[41–43] However, cation-driven hydrogels of G-C18:1 display an impressive structural stability to shear and mechanical resistance to temperature. {Ag$^+$}G-C18:1, more stable than {Ca$^{2+}$}G-C18:1 hydrogels, keeps both their structure and elastic properties up to 70°C, whereas the melting temperature of the C18:1 backbone is known to be below room temperature. Last but not least, G-C18:1 is obtained by fermentation from natural resources, demonstrating biobased amphiphiles can have unique properties, thus possibly replacing synthetic amphiphiles in the long run for a more sustainable future.

**Material and methods**

*Chemicals.* The monounsaturated glucolipid G-C18:1 ($M_w$ = 460 g.mol$^{-1}$) contains a β-D-glucose unit covalently linked to oleic acid (Figure 1a). The molecule is obtained by fermentation from the yeast *Starmerella bombicola ΔugtB1* according the protocol given before.[21,44] The compound is purchased from the Bio Base Europe Pilot Plant, Gent, Belgium, lot N° APS F06/F07, Inv96/98/99 and used as such. According to the specification sheet provided by the producer, the batch (99.4% dry matter) is composed of 99.5% of G-C18:1, according to HPLC-ELSD chromatography data. NMR analysis of the same compound (different batch) was performed elsewhere.[20] NaOH (≥98wt% pellets) is purchased from Sigma Aldrich, CaCl$_2$ in pellets and liquid 35wt% HCl are purchased from VWR. AgNO$_3$ is purchased from Sigma Aldrich.

*Sample preparation.* G-C18:1 is dispersed in milli-Q water and the pH is adjusted by an initial addition of concentrated NaOH (5 M), followed by a refinement with few µL of more diluted NaOH (or HCl) solution (1 M, 0.5 M or 0.1 M). The targeted molecular ratio to be roughly in the region of pH 8 is [NaOH]/[G-C18:1]= 0.7–0.8. The solution is homogenized by vortexing. To form a hydrogel, a CaCl$_2$, or AgNO$_3$, solutions are prepared at 1 M and the appropriate amount is added to the G-C18:1 solution according to the molar ratio [AgNO$_3$]/[G-C18:1] = 1.0 and [CaCl$_2$]/[G-C18:1] = 0.6. Typically, for a total 1 mL volume, 62.5 µL of AgNO$_3$ (1 M), or 40 µL of CaCl$_2$ (1 M), are added to the complementary volume of 3 wt% G-C18:1. After the addition, the solution is immediately stirred during about 30 s. All samples are aged between few minutes and 3 days before any experiments. Hydrogels prepared from Ca$^{2+}$ and Ag$^+$ are respectively labelled {Ca$^{2+}$}G-C18:1 and {Ag$^+$}G-C18:1.

Throughout the text, the salt-to-lipid ratios are provided either in the form cation-to-G-C18:1 ratio, $[M^{z+}]/[G\text{-}C18:1]$, with $M^{z+} = Ca^{2+}$ and $Ag^+$, or, when relevant, in the form positive-to-negative charge ratio, $[X^+]/[G\text{-}C18:1]$, with $[X^+] = 2[Ca^{2+}]$ or $[Ag^+]$, considering that at pH> 8, all G-C18:1 molecules bear a negative charge related to the ionized carboxylic acid.

*Rheology.* A MCR 302 rheometer (Anton Paar, Graz, Austria) is used with sand-blasted plate-plate geometry (Ø: 25mm, gap between 0.5 – 1 mm) and a cone-plate geometry (Ø: 50mm) at a regulated temperature of 25°C. Solvent trap with water is used to minimize evaporation. $\sim 0.5$mL of gel is loaded on the center of the plate using a spatula to prevent trapped bubbles, then the excess is removed. Value of the pseudo-equilibrium G' is taken after 5 min of oscillatory measurement at 1 Hz and low strain $\gamma$, one order of magnitude lower than the critical strain.

*Small angle X-ray scattering (SAXS).* SAXS experiments have been performed on various beamlines and synchrotron facilities. The environments as well as the samples associated to each session are presented below.

*Capillary SAXS.* SAXS experiments are performed using hydrogel samples prepared at room temperature and analyzed into 1.0 mm quartz capillaries on the BM29 beamline (Proposal N° MX 2311) at the ESRF Synchrotron (Grenoble, France). Samples are manually injected into the capillary using a 1.0 mL syringe. The BM29 beamline is used with an energy of E = 12.5 KeV and a sample-to-detector distance of 2.83 m. q is the wave vector, with $q = 4\pi/\lambda \sin(\theta)$, $2\theta$ corresponding to the scattering angle and $\lambda$ the wavelength. The q-range is calibrated between $\sim 0.05 < q \,/\, nm^{-1} < \sim 5$, using the standard silver behenate calibrant ($d_{(001)} = 58.38$ Å); raw data obtained on the 2D detector are integrated azimuthally using the in-house software provided at the beamline and thus to obtain the typical scattered intensity I(q) profile. Absolute intensity units are determined by measuring the scattering signal of water (I(q=0) = 0.0163 $cm^{-1}$).

*Rheo-SAXS.* The first set of data is collected on the ID02 beamline at the ESRF-EBS[45] synchrotron (Grenoble, France) during the proposal N° SC-4976. The energy of the beam is set at 12.28 keV and the sample-detector distance at 1.5 m. The beamline is equipped with a Haake Rheo-Stress RS6000 stress-controlled rheometer containing a polycarbonate couette cell having a gap of 0.5 mm and a required sample volume of 3 mL. The rheometer is controlled through an external computer in the experimental hutch using the control software RheoWin. The temperature of the cell is set at 25°C, unless otherwise stated. Experiments are performed in a radial configuration i.e. the X-ray beam is aligned along the center of the Couette cell. The absolute intensity I(q) is obtained by standard normalization and subtracting the background (polycarbonate cell containing milliQ water) and by dividing the signal by the effective thickness of the sample in cell (0.17 cm). The SAXS acquisitions are manually triggered at the same time as the rheology acquisition, with an error of $\pm 2$ s. The frequency of data recording is manually set and determined independently for both SAXS and rheology measurement. Shear and oscillatory measurement are performed at various velocity, strain and frequency.

The second set of data is recorded at the SWING beamline of the Soleil synchrotron facility (Saint-Aubin, France) during the run N° 20200532, using a beam energy of 12.00 keV and a sample-detector distance of 1.65 m. Tetradecanol ($d_{(001)}$ = 39.77 Å) is used as the q-calibration standard. The signal of the EIGERX 4M 2D detector (75 μm pixel size) is integrated azimuthally with Foxtrot software to obtain the I(q) spectrum (q =$4\pi \sin \theta/\lambda$, where $2\theta$ is the scattering angle) after masking systematically defective pixels and the beam stop shadow. A MCR 501 rheometer (Anton Paar, Graz, Austria) equipped with a Couette polycarbonate cell (gap 0.5 mm, V= 1.35 mL) is coupled to the beamline and controlled through an external computer in the experimental hutch using the Rheoplus/32 V3.62 software. The experiments are performed in a radial configuration, where the X-ray beam is aligned along the center of the Couette cell. The rheology and SAXS acquisitions are synchronized manually with an estimated time error of less than 5 s. Due to standard security procedures, the first rheo-SAXS experimental point is systematically acquired with a delay of about 2–3 minutes with respect to the rheometer. Data are not scaled to absolute intensity. Rheo-SAXS samples each require a $\approx$ 2 mL volume.

| Figure | Sample | Experiment | Beamline | Synchrotron | Proposal N° |
|---|---|---|---|---|---|
| 1 | All | Capillary SAXS | BM29 | ESRF | MX 2311 |
| 3a, S 3 | All | | | | |
| 3c | pH 8 | | | | |
| 3c | pH 6.5, pH 7 | Rheo-SAXS | ID02 | ESRF | SC-4976 |
| 4a, 4b, S 4 | All | | | | |
| 4c, 4d | All | Rheo-SAXS | SWING | Soleil | 20200532 |

*Dynamic Scanning Calorimetry (DSC).* DSC is performed using a DSC Q20 apparatus from TA Instruments equipped with the Advantage for Q Series Version acquisition software (v5.4.0). Acquisition is performed on: dry G-C18:1 powder (15.3 mg), freeze-dried {Ca²⁺}G-C18:1 fibers (15.3 mg), freeze-dried {Ag$^+$}G-C18:1 fibers (15.3 mg), a {Ca$^{2+}$}G-C18:1 fiber fiber aqueous solution at C= 3 wt%, a partially hydrated {Ca$^{2+}$}G-C18:1 fiber sample (uncontrolled hydration). All samples are sealed in a classical aluminium cup and using an immediate sequence of heating and cooling ramps at a rate of 2°C.min$^{-1}$. Melting temperatures, $T_m$, 1 and 2, $T_{m1}$ and $T_{m2}$, are taken at the minimum of the endothermic peak.

*Cryogenic-transmission electron microscopy (Cryo-TEM).* Pictures are recorded on an FEI Tecnai 120 twin microscope operating at 120 kV with an Orius 1000 CCD numeric camera. The sample holder is a Gatan Cryo holder (Gatan 626DH, Gatan). Digital Micrograph software is used for image acquisition. Cryo-fixation is done with low dose on a homemade cryo-fixation device. The solutions are deposited on a glow-discharged holey carbon coated TEM nikel grid (Quantifoil R2/2, Germany). Excess solution is removed and the grid is immediately plunged into liquid ethane at -180°C before transferring them into liquid nitrogen. All grids are kept at liquid nitrogen temperature throughout all experimentation. Cryo-TEM images are treated and analyzed using Fiji software, available free of charge at the developer's web site.[46]

## Results

Bolaform glucolipid G-C18:1 contains a free-standing carboxylic acid end-group on its alkyl chain (Figure 1a). It behaves as a surfactant above and as a lipid below neutrality of water (pH ~7) at room temperature: a micellar phase is observed at basic pH while a vesicular phase is reported at acidic pH.[20,21] Qualitative experiments where various sources of cations are added below neutrality, in the vesicle phase, show precipitation, most likely of a lamellar aggregate, as reported for other metal-vesicle systems[47] and in analogy to what it was reported before for the same compound at pH below ~4.5.[20,21] However, adding alkaline earth and transition metal cations in the micellar phase above neutrality induces gelling,[22] whereas NaCl, which does not perturb the micellar phase up to molar amount, acts as a negative control.[22,37] The effect of cations on both the structure and the elastic properties of the hydrogels is reported elsewhere,[22] while here we focus on the specific effect of calcium and silver ions, of which the derived hydrogels share similar elastic and structural features.

Specific structural[37] and mechanistic[48] studies have performed on {Ca$^{2+}$}G-C18:1 and {Ag$^+$}G-C18:1 hydrogels and reported elsewhere, in parallel to this work.[37,48] Both {Ca$^{2+}$}G-C18:1 and {Ag$^+$}G-C18:1 form hydrogels composed of self-assembled fibers, characterized by the uncommon coexistence of entangled fibers and β-sheet-like fibrillar raft domains. Typical cryo-TEM data recorded on {$Ca^{2+}$}G-C18:1 and {$Ag^+$}G-C18:1 hydrogels, reported in Figure 1b, show that the fibrillar network is composed of domains, which are constituted by side-by-side association of fibers in β-sheet-like rafts, pointed at by white arrows 1-3 for $Ca^{2+}$ and 4-6 for $Ag^+$. Highlights of arrows 1 and 5-6 are also provided on the same figure. Although not a quantitative technique, cryo-TEM shows that calcium-based samples rather have a multimeric association of fibers, while silver-based samples are more characterized by dimers and trimers. Cryo-TEM data are corroborated by corresponding, statistically-relevant, SAXS experiments. Typical SAXS of {$Ca^{2+}$}G-C18:1 and {$Ag^+$}G-C18:1 hydrogels are shown in Figure 1c. In the absence of added salt, G-C18:1 displays a scattering profile (black) typical of micelles in water ($q > 0.2$ nm$^{-1}$) coexisting with larger structure ($q < 0.2$ nm$^{-1}$). When either $Ca^{2+}$ or $Ag^+$ ions are added to the micellar solution, the solution immediately forms a gel and the corresponding SAXS displays a low-q scattering profile having a slope close to -2 (log-log scale), a value generally associated to flat structures, including nanotapes and ribbons.[49] The SAXS curves of both {$Ca^{2+}$}G-C18:1 and {$Ag^+$}G-C18:1 hydrogels also show a series of diffraction peaks, of which the q-values have an integer multiplicity (1:2:3…) up to the fifth order, typical of well-ordered lamellar structures. The fibrillar nature of the hydrogels shown by cryo-TEM excludes the presence of liquid crystalline lamellar phases, but it corroborates the hypothesis of the fibrillar rafts. A more refined structural analysis of {$Ca^{2+}$}G-C18:1 and {$Ag^+$}G-C18:1 hydrogels as well as the role of the metal-ligand interaction are given in ref.[37].

The most appropriate way to interpret the cryo-TEM and SAXS data related to {$Ca^{2+}$}G-C18:1 and {$Ag^+$}G-C18:1 hydrogels is not the classical fiber entanglements,[50–52] or possible Van der Waals cross-linking,[53] but rather the so-called "nano-fishnet" description (Figure 1d), known for silk fibroin or actin proteins gels.[12–18] Evidence of β-sheet-like rafts in amphiphilic and small peptidic fibrillar gelled systems is rare, or even non-existing, as summarized and commented in the broad literature survey in Table S 1. As proposed for proteins, the driving force to obtain β-sheet-like rafts with G-C18:1 could be explained by possible multipolar moments, as intended by Van Workum et al.[54] However, the unique molecular structure of G-C18:1 compared to similar compounds and additional $^{13}C$ solid state NMR arguments tend to identify specific metal-ligand interactions as the main driving force.[19]

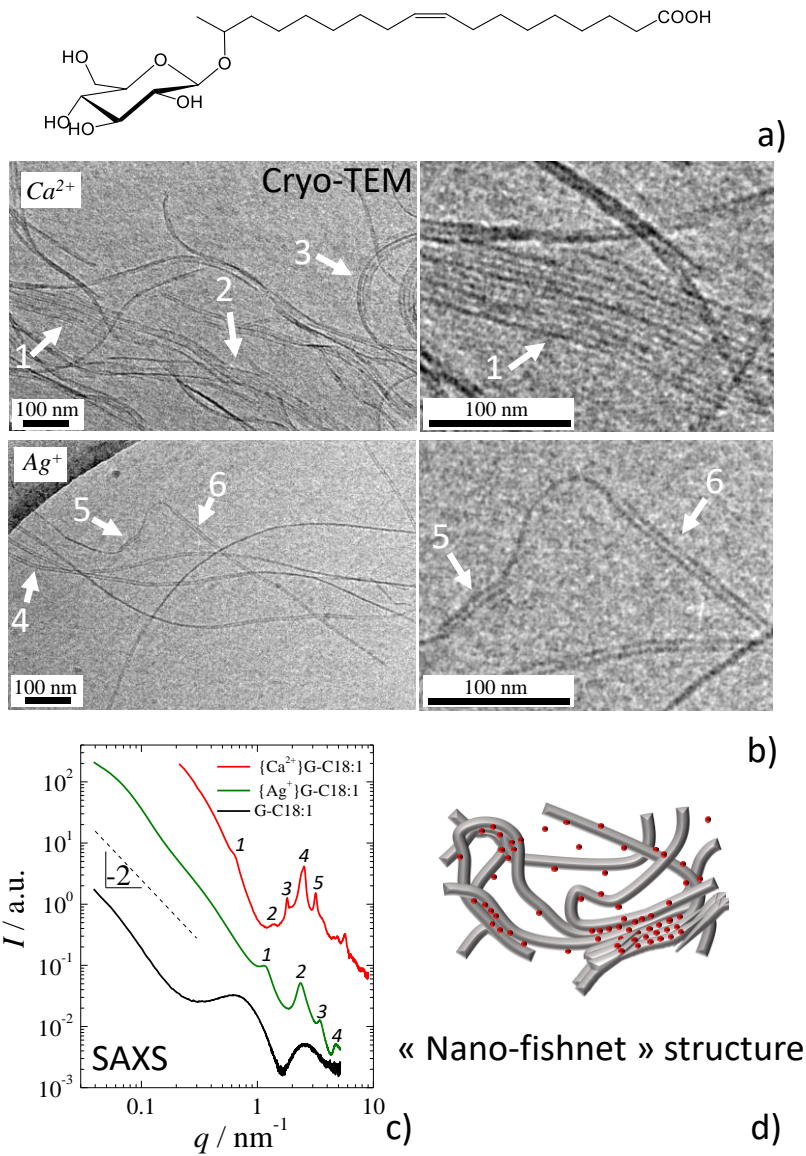

**Figure 1 – a) Molecular formula of G-C18:1. Structural characterization of {Ca²⁺}G-C18:1 and {Ag⁺}G-C18:1 hydrogels: b) cryo-TEM images of 0.5 wt% G-C18:1 at a ratio [Ca²⁺]/[G-C18:1] = 0.6, [Ag⁺]/[G-C18:1] = 1 in water at a basic pH. Samples are not sheared. c) SAXS profiles of 2 wt% of G-C18:1 in water at a basic pH in the absence (black) and presence of Ca²⁺ (red) and Ag⁺ (green). d) Representative model of the "nano-fishnet" structure.**

Hereafter, we report the in-depth characterization of the elastic properties of {Ca²⁺}G-C18:1 and {Ag⁺}G-C18:1 hydrogels having a "nano-fishnet" structure.

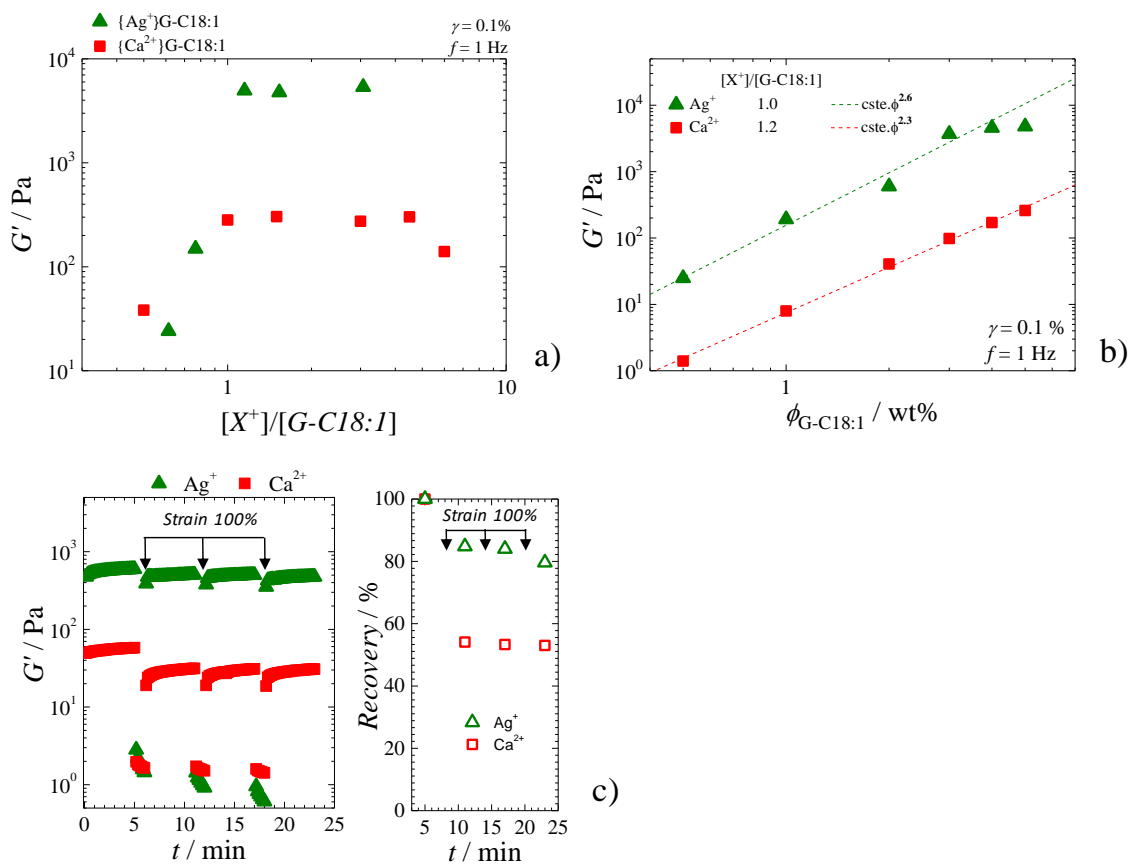

**Figure 2 -** Elastic moduli, G', of {Ca²⁺}G-C18:1 (square) and {Ag⁺}G-C18:1 (triangle) hydrogels at basic pH as a function of : a) charge-normalized cation-lipid ratio (C_G-C18:1= 3 wt%) and b) G-C18:1 weight fraction, φ_G-C18:1. The dash lines correspond to a linear power law fitting. c) Step-strain experiment (γ= 0.1 %, 100%, f= 1 Hz). Recovery is calculated with respect to the plateau between 0 and 5 min. In b) and c), the charge ratio is [X⁺]/[G-C18:1]= 1.2 and 1 for Ca²⁺ and Ag⁺, respectively, where [X⁺] represents the charge-normalized cation molar concentration, [X⁺]≡ [Ag⁺] and [X⁺]≡ 2[Ca²⁺]. For instance, [X⁺]/[G-C18:1]= 1.2 is equivalent to [Ca²⁺]/[G-C18:1]= 0.6. In a) and c), the G-C18:1 concentration, C_G-C18:1= 3 wt%.

The linear viscoelastic regime of both {Ca²⁺}G-C18:1 and {Ag⁺}G-C18:1 is determined by a strain sweep experiment, while the gel properties (G' > G'') are confirmed by frequency sweep experiments at selected [X⁺]/[G-C18:1] ratios (Figure S 1). [X⁺] is the molar concentration of the positive charges, taken as 2[Ca²⁺] and [Ag⁺] for Ca²⁺ or Ag⁺, respectively. The charge ratio dependency of the elastic modulus (Figure 2a) shows the formation of a pseudo-plateau above about a charge ratio of [X⁺]/[G-C18:1]= 1.05 (C_G-C18:1= 3 wt%) that is 68 mM in the case of silver hydrogels (triangle). The average G' is about 3.7 kPa (C_G-C18:1= 3 wt%) but values as high as about 5 kPa can be recorded. In the case of calcium hydrogels (square), the elastic properties reach a plateau of G' ~ 0.25 kPa at already a charge ratio of 1.2, reflecting a concentration ratio [Ca²⁺]/[G-C18:1]= 0.61 and [Ca²⁺]= 40 mM.

Plotting G′ against the G-C18:1 weight fraction, $\varphi_{G-C18:1}$, at a fixed [ion]/[G-C18:1] ratio corresponding to the plateau ([X+]/[G-C18:1]> 1) shows that silver hydrogels are at least one order of magnitude stronger than calcium gels (Figure 2b). G′ scales with $\varphi_{G-C18:1}$ following a power law of 2.3 and 2.6 for calcium and silver hydrogels, respectively. Scaling law with a power dependency between 2.3 and 2.6 are typical for polymers and fibrils in a good solvent.[55–57]

The improved strength, but also recovery rate of {Ag+}G-C18:1 gels is also measurable by step-strain experiments within 5 min time frame (Figure 2c), considered to be enough when compared to longer time scales (Figure S 2). If both {Ca2+}G-C18:1 and {Ag+}G-C18:1 gels recover rapidly after applying a 100% strain, {Ag+}G-C18:1 gels show a recovery of more than 80%, compared to only 55% of {Ca2+}G-C18:1 gels, after 30 s from strain release. The lower recovery rate of calcium-based hydrogel could be correlated with the slower formation of β-sheet-like domains, more sensitive to shear in {Ca2+}G-C18:1 than in {Ag+}G-C18:1 systems, as observed by *ex situ* and *in situ* SAXS.[37,48]

*pH effects*

Calcium or silver hydrogels only form in the micellar region above pH ~7. This phenomenon is against the common belief that surfactant micelles elongate into cylindrical, or wormlike,[26–29,32] objects (more rarely, vesicular or lamellar[29,31]) in the presence of salts. As a first observation, gels of slightly higher strength and of better defined crystallinity are formed at pH 10, rather than pH 8 (Figure 3a,b), independently of the glycolipid concentration (Figure S 3). If the initial pH does not have a dramatic impact on the final structure and properties, qualitative observations suggest faster gelation at pH 10 than at pH 8 for a given G-C18:1 concentration and [ion]/[G-C18:1] molar ratio.

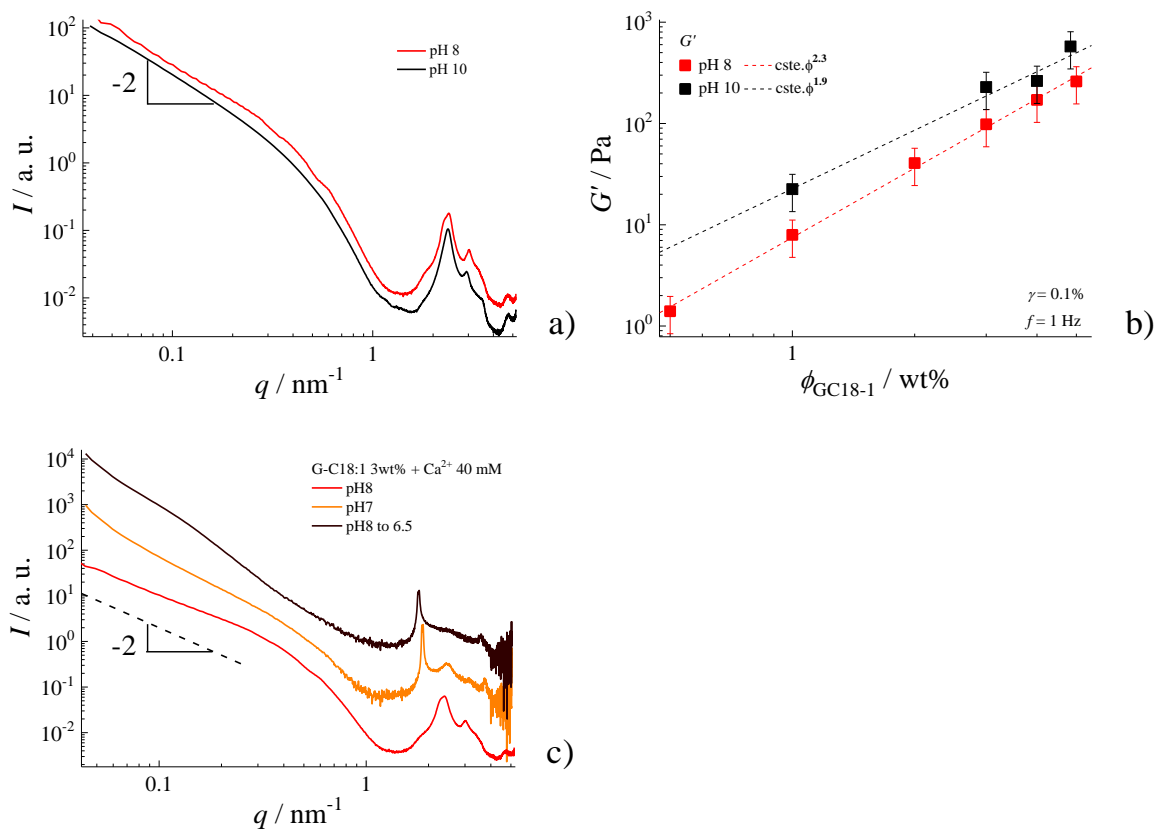

**Figure 3 - a) SAXS spectra and b) gel strength of a {Ca²⁺}G-C18:1 gel ([Ca²⁺]/[G-C18:1]= 0.61) at pH 8 (red) and 10 (black). c) SAXS spectra of {Ca²⁺}G-C18:1 gel ([Ca²⁺]/[G-C18:1]= 0.61 and G-C18:1= 3 wt%) prepared at pH 8, pH 7 and after an acidification from pH 8 to 6.5 (curves are shifted by a factor 10).**

On the other hand, combining $Ca^{2+}$, or $Ag^+$, with acidification induces massive precipitation into a powder with a lamellar structure at pH close to neutrality, as shown by the variation in the diffraction peaks pattern recorded by SAXS from pH 8 to 6.5 (Figure 3c) for calcium based gels. The effect of adding an excess of ions in the vesicle phase generates aggregates, as found elsewhere,[47] and it has in fact a similar effect as reducing pH below 4.[20,21]

The pseudo plateau of G' as a function of ion concentration (Figure 2a) is reached at a [X⁺]/[G-C18:1] molar ratio of about 1 for $Ag^+$ and $Ca^{2+}$ ([X⁺] is the normalized charge molar cation concentration, as explained in the legend of Figure 2), indicating that the optimum elastic properties are related to the positive/negative charge balance. Complementary ion-resolved *in situ* SAXS studies performed on diluted {Ca²⁺}G-C18:1 solutions provide additional pieces of information:[48] 1) fibers form in the range between 0.2 < [X⁺]/[G-C18:1] < 0.6 (corresponding to 0.1 < [Ca²⁺]/[G-C18:1] < 0.3); 2) crystallization starts at [X⁺]/[G-C18:1] around 1.2-1.4 (corresponding to [Ca²⁺]/[G-C18:1] > 0.6-0.7). Supplementary *in situ* SAXS data[48] combined with rheology experiments performed here (Figure 2a) suggest then that optimum gelling occurs upon crystallization of the fibers and their consequent assembly into lamellar rafts, the latter commented in more detail in Ref. [37]. Previous structural studies attribute crystallization to the coordination of the carboxylate group of G-C18:1 by $Ca^{2+}$ and eventual rearrangement of G-C18:1 within the fibers. [37]

*Stability against temperature*

SAFiN gels are characterized by a gel-to-sol transition temperature,[58] often corresponding to a fiber-to-micelle phase transition. Ion-induced fibrillation of G-C18:1 occurs at room temperature, thus suggesting that cations modify the temperature behavior of this specific lipid. Their temperature stability could then be intriguing, and in particular if related to the nano-fishnet hydrogel structure.[37] Rheo-SAXS experiments offer the possibility to study the macroscopic viscoelasticity in association to the fiber's structure during a temperature ramp.

In Figure 4, the elastic and viscous moduli of $Ag^+$ (circle) and $Ca^{2+}$ (square) hydrogels prepared on the charge ratio plateau (Figure 2a) are studied by rheo-SAXS as a function of temperature. At room temperature, the elastic modulus of {$Ag^+$}G-C18:1 hydrogels is only 150 Pa. This lower value compared to the ones in Figure 2 under similar conditions is explained by the nature of the couette cell sample holder on the SAXS beamline and known to apply an initial stronger stress than a plate-plate geometry. However, this modulus is still sensibly higher than for the {$Ca^{2+}$}G-C18:1 hydrogel, in agreement with the study on plate-plate geometry. One should note that the characteristic features of the SAXS profiles in Figure 4b,d will not be discussed in this work, but they can be found in ref. [37].

Upon heating, at 70°C during 15 min, and after cooling, the {$Ag^+$}G-C18:1 gel is astonishingly stable, both in terms of its practically constant viscoelastic moduli (Figure 4a) and microstructure (unchanged SAXS profiles in Figure 4b). The elastic behaviour of the {$Ca^{2+}$}G-C18:1 hydrogel remains stable from room temperature until about 60°C. Increasing temperature up to 63°C induces a gel-to-sol transition (Figure 4a), although still characterized by a SAXS signal typical of a fibrillar structure (Figure 4b). From about 63°C and up to 70°C, in the liquid state, the SAXS shows a progressive fiber-to-micelle phase transition (Figure 4b and Figure S 4), where micelles have a gyration radius, $R_g$= 3.18 nm, consistent with the typical length of G-C18:1.[20,21] The sol-to-gel transition progressively takes place again upon cooling and in direct relationship to the formation of the fiber phase (SAXS, Figure 4b).

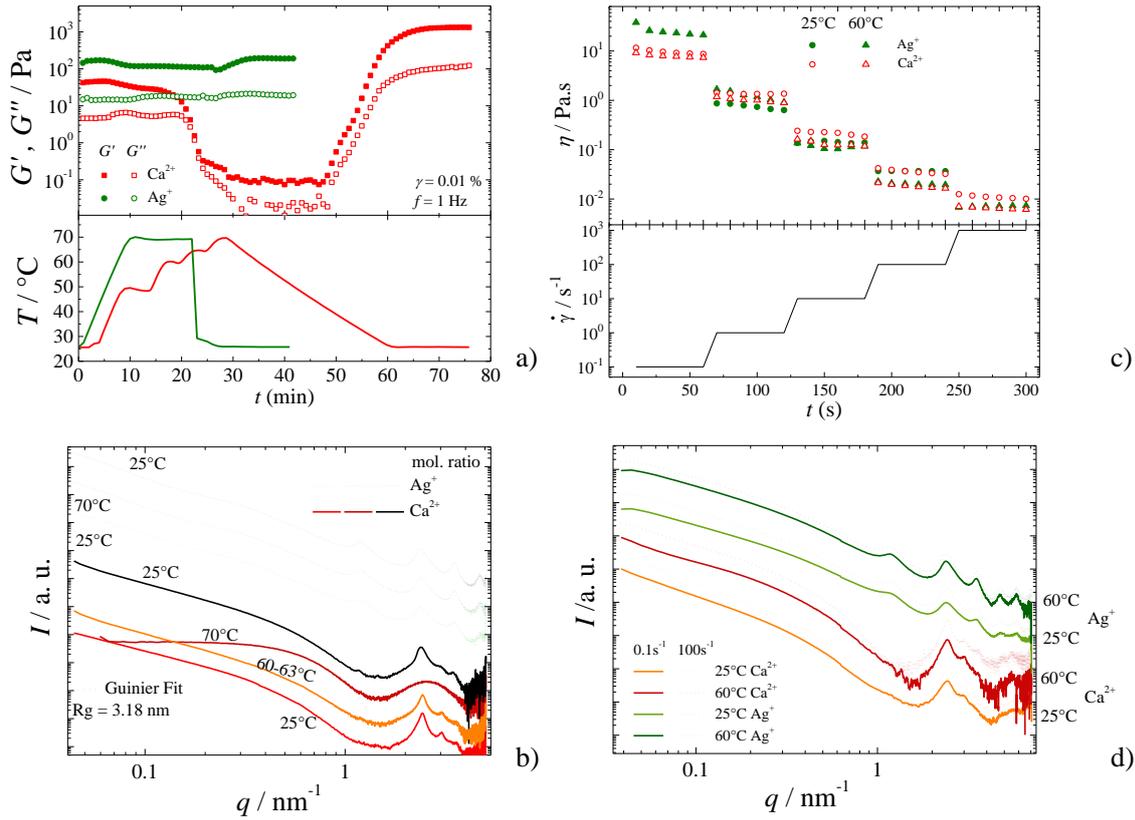

**Figure 4 - Rheo-SAXS experiment. a) Oscillation-temperature profile and b) SAXS spectra corresponding to 3 wt% of G-C18:1 at basic pH with a [Ag+]/[G-C18:1]= 1.0 (circles in (a), dash line in (b)) and [Ca²⁺]/[G-C18:1]= 0.61 (square in (a), solid line in (b)). Dotted black line in b) corresponds to a Guinier fit. c) Shear-temperature profile and d) corresponding SAXS spectra for a 1 wt% of G-C18:1 at a [Ca²⁺]/[G-C18:1]= 0.61 (empty symbol) and [Ag+]/[G-C18:1]= 1.0 (solid symbol) at basic pH. Experiments are performed both at 25°C and 60°C. In b) and d), data are artificially shifted. Shift factors are arbitrarily chosen.**

The stability against temperature has also been studied under dynamic conditions. Figure 4c,d correspond to rheo-SAXS experiment under rotation. The dynamic viscosity, η, decreases with increasing shear rate, γ̇, for all samples (Figure 4c), illustrating the typical phenomenon of rheo-thinning in SAFiN hydrogels. However, the most interesting aspect is related to the stability of all hydrogels at a given γ̇, irrespective of the temperature: η is practically constant for {Ca²⁺}G-C18:1 and {Ag+}G-C18:1 hydrogels at 25°C (circles, Figure 4c) and 60°C (triangles, Figure 4c) for a well-defined γ̇, in the wide range $0 < \dot{\gamma} \, / \, s^{-1} < 1000$. Interestingly, the loss in viscosity is not correlated to a loss in structure, as the corresponding SAXS profiles are not sensibly altered between 0.1 s⁻¹ and 100 s⁻¹ both at 25°C and 60°C (Figure 4d). Oscillatory and rotational rheo-SAXS show the stability of {Ca²⁺}G-C18:1 and {Ag+}G-C18:1 hydrogels with temperature, and, for the specific case of Ca²⁺, they show that the gel-to-sol transition is controlled by a network collapse prior to the fiber-to-micelle phase transition. This is unexpected and it could corroborate the hypothesis according to which both entanglement and β-sheet-like rafts[37] contribute to stabilize the G-C18:1 hydrogels.

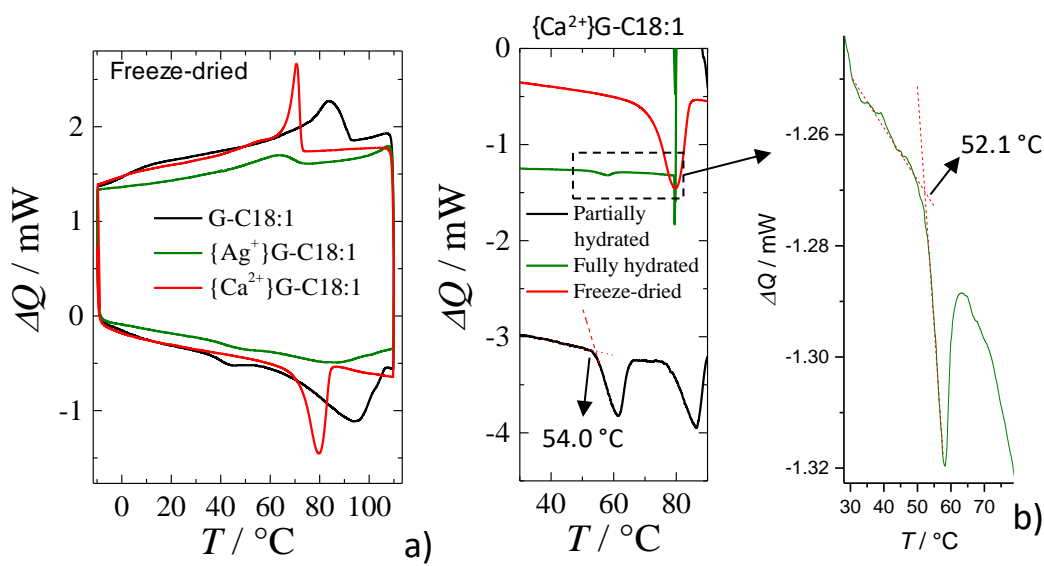

Figure 5 - DSC thermograms (negative peaks are endothermic) of a) G-C18:1 powder control and {Ca²⁺}G-C18:1, {Ag⁺}G-C18:1 freeze-dried fibers; b) freeze-dried, partially (uncontrolled) hydrated, fully hydrated (C$_{H2O}$= 97 wt%) {Ca²⁺}G-C18:1 samples.

**Table 1 – Data from DSC analysis presented in Figure 5**

| Sample | State | Tm$_1$ / °C | Tm$_2$ / °C |
|---|---|---|---|
| G-C18:1 | Freeze-dried | 37.6 | 68.1 |
| | | | |
| {Ca²⁺}G-C18:1 | Freeze-dried | - | 72 |
| {Ca²⁺}G-C18:1 | Partially hydrated | 54.9 | 78.3 |
| {Ca²⁺}G-C18:1 | Fully hydrated | 52.1 | - |
| | | | |
| {Ag⁺}G-C18:1 | Feeze-dried | 36.7 | |

Dynamic scanning calorimetry (DSC, Figure 5a) correlates well with such a picture. Salt-free G-C18:1 (powder) is characterized by a melting temperature, Tm, at 68.1 °C and a pre-melting transition at 37.6 °C (Table 1). They certainly find their origin in the internal phases transitions of the alkyl chain as observed for a saturated glucolipid of 18 carbons atoms without the carboxylic group, having two phases transition at 56.7°C and 72.5°C.[59] {Ca²⁺}G-C18:1 freeze-dried fibers have a sharp Tm, slightly shifted at 72°C, and no obvious pre-transition. However, upon partial (uncontrolled) and full (97 wt% H$_2$O) hydration, a well-defined endothermic peak appears between 52°C and 55°C (Figure 5b), and which could explain the network collapse in the gel-to-sol transition prior to the lipid melting. In the case of the silver-based freeze-dried fibrillar sample, the endothermic transitions are much weaker, and the corresponding temperatures are curiously comparable to the salt-free glucolipid. This probably suggests that residual amount of uncomplexed G-C18:1 could still part of the sample, while the melting of {Ag$^+$}G-C18:1 itself occurs at temperatures higher than the ones explored here.

*Structure-properties relationship*

G-C18:1 with Ag$^+$ or Ca$^{2+}$ forms gels, of which the network elasticity is characterized by physical junctions. The elastic properties of most SAFiN hydrogels reported in the literature are related to fibers' entanglement.[9,14,41] On the other hand, those systems showing spontaneous bundling do not necessarily form hydrogels (Table S 1). The elastic properties and stability of {Ca$^{2+}$}G-C18:1 and {Ag$^+$}G-C18:1 hydrogels seem to depend on entanglement, as supposed from the scale law dependency of G'($\varphi$) being close to 2.25. However, it is not excluded that the high elastic moduli and resistance to shear and temperature also depend on junctions constituted by the β-sheet-like rafts (and ultimately, metal-ligand interactions), characterized elsewhere.[37] Several evidences suggest this fact.

Shear generally induces orientation in both fibrillar and lamellar systems. However, up to at least 100 s$^{-1}$, 2D images recorded by rheo-SAXS experiments show an isotropic, not oriented, signal associated to the crystalline raft domains (Figure S 5). This result is quite unexpected for fibrous systems and it strengthens the idea of the "nano-fishnet" structure,[37] composed of isotropically-distributed rafts.

 In the meantime, hydrated samples show a well-defined endothermic peak about 13°C to 16°C below the Tm of G-C18:1. Such peak can be reasonably associated to the gel-to-sol transition probed by rheology, while the fiber-to-micelle phase transition probed by SAXS above 63°C is rather associated to the melting of the lipid, represented by the second endothermic peak above 65°C in DSC.

The lamellar signature of the rafts[37] and elastic properties of the gels depend on the type of cation and sample history. Sheared {Ca$^{2+}$}G-C18:1 gels display almost no β-sheet-like structures, and their elastic properties are moderate. On the other hand, {Ag$^+$}G-C18:1 gels form thinner fibers constituted by dimeric G-C18:1 complexes bridged by silver ions. Fibers associate into less wide, but more stable, β-sheet-like rafts.[37] Silver is more reactive in contact with an alkaline G-C18:1 solution, as shown by ITC,[37] and gels form faster,[48] with enhanced mechanical strength (Figure 2).

A common way to estimate the rigid, or freely, hinged behaviour of junctions is modelling the power law dependence of G' as a function of the gelator volume fraction ($\varphi_{\text{G-C18:1}}$, Figure 2). According to the theory developed by Jones and Marques, the entropic or enthalpic nature of the junction is associated to the fractal dimension, $D_f$, of the fibril.[60] $D_f$ is related to the ability for the fibril to maximize/minimize its contact with the solvent, e.g., $D_f=$ 1 for a good solvent (infinitely extended rod), $D_f=$ 2 for a bad solvent (mass fractal). Enthalpic and entropic elasticity refer to the degree of freedom, intended as variability of the interfibril angle at the junctions, being respectively equal or different than zero. Enthalpic elasticity then attributes the reduction in degree of freedom to the rigidity of the crosslinks and fibrils. For entropic elasticity, fibrils are freely hinged and it is the high number of fibrils per crosslink, which reduce the degree of freedom of the system. In both cases, junctions are considered permanent and at a fixed point of the fibril.[61]

To apply the model to {$Ca^{2+}$}G-C18:1 and {$Ag^+$}G-C18:1 hydrogels, we assume the density of G-C18:1 close to 1, meaning that weight and volume fractions are equivalent, and we also assume a negligible G-C18:1 fraction out of the elastic network. Considering that in the linear viscoelastic regime, the Young modulus, E, is equivalent to G', one can write $E \propto \varphi^\alpha$, whereas

$$E_{entropic} = k_b T \varphi^{\left(\frac{3}{3-D_f}\right)}, \alpha = \frac{3}{3-D_f}$$

$$E_{enthalpic} = \varphi^{\left(\frac{3+D_f}{3-D_f}\right)}, \alpha = \frac{3+D_f}{3-D_f}$$

with $k_b$ being the Boltzmann constant, T the temperature and $\varphi$ the volume fraction of elastic materials. From Figure 2, the experimental $\alpha$ are $\alpha_{\text{exp}}=$ 2.3 for {$Ca^{2+}$}G-C18:1 and $\alpha_{\text{exp}}=$ 2.6 for {$Ag^+$}G-C18:1 gels. According to the cryo-TEM analysis,[37] fibers are well-dispersed in the medium, thus being better described by $D_f=$ 1. The calculated $\alpha$ are then $\alpha_{\text{calc}}=$ 1.5 and 2.0 respectively for the entropic and enthalpic contributions. Comparison between $\alpha_{\text{exp}}$ and $\alpha_{\text{calc}}$ suggests that the enthalpic contribution prevails, in agreement with rigid cross-links, as proposed above on the basis of cryo-TEM and SAXS experiments.

Gelation by silver occurs at lower $\varphi_{\text{G-C18:1}}$ and G' is at least ten times higher than with calcium (Figure 2). This could be explained by the better stability of the carboxylate-silver coordination, as suggested by the highly exothermic heat exchange observed by ITC,[37] and indicating an efficient and more stable bridging, rather than a charge neutralization effect. In terms of comparison, $Na^+$ does not induce fibrillation and the heat exchange profile of {$Ca^{2+}$}G-C18:1 is characterized by non-specific endothermic and specific exothermic reactions, respectively, with a less negative $\Delta H$.[37] $Ag^+$ is known to form a crystalline structure with a large number of fatty acids by complexation with their $COO^-$.[62,63]

$Ca^{2+}$ and $Ag^+$ have at least two contributions:[37] they induce a micelle-to-fiber transition, stabilizing the latter, and they most likely act as bridge-like cross-linkers, but only on portions of the fibers, thus generating multimer rafts on length scales from ~10 nm to several hundred nm, but not precipitation of large crystals. This could only be explained by the inhomogeneous adsorption of ions at the fibers' surface, and in particular to the side edges. Inhomogeneous adsorption of ions onto lipidic membranes is a classical phenomenon observed for flat and curves lipid membranes.[64] However, to the best of our knowledge, raft-like structures of fibers are extremely rare, if not unique, for colloidal SAFiN solutions.[40] They are more reminiscent of imogolite (mineral aluminosilicates) nanotube solutions or specific protein hydrogels, like silk fibroin or actin, where network elasticity is related to β-sheet junctions.[12–18] In this regard, we have proposed a specific dimeric building unit within the fibers,[37] of which the core is rich in cations and the external sides in glucose, similarly to what has been reported for silver stearate.[63] Interestingly, direct proof of the metal-ligand interaction is provided by $^{13}C$ solid-state nuclear magnetic resonance (ssNMR) experiments recorded on freeze-dried micellar and gel solutions of G-C18:1.[19] ssNMR not only shows different peak width and chemical shift between the micellar and fiber samples, but it also shows a clear-cut splitting of the $^{13}C$ resonance attributed to the $COO^-$ group for $\{Ca^{2+}\}$G-C18:1 gels, thus supporting the hypothesis of two binding sites for $Ca^{2+}$, the first within each fiber and the second across fibers.[19]

Differently than fatty acid-metal complexes, G-C18:1 fibers are colloidally stable in water. This phenomenon can only be explained by the presence of the glucose headgroup, which makes the molecule double hydrophilic. Glucose could also have another role. Its bulkiness could induce flip flops in the packing of G-C18:1, so to maximize packing and minimize steric repulsion between adjacent glucose groups. Flip flops could induce structural defects in the fiber's structure, thus explaining an accumulation of negative charges on the external sides of the fibers and responsible for bridging between adjacent fibers.[37] This is specifically observed for G-C18:1, as similar glycolipid, only differencing in the number (sophorolipids) or type (rhamnolipids) of sugar unit, do not show any specific morphological change when exposed to a source of monovalent or divalent cations.[23,25]

Due to this particular structure and ion complexation, both $\{Ca^{2+}\}$G-C18:1 and $\{Ag^+\}$G-C18:1 hydrogels display unexpected resistance to both temperature and shear, up to at least 65°C for $\{Ca^{2+}\}$G-C18:1 and above 80°C for $\{Ag^+\}$G-C18:1 hydrogels. For instance, other fibrillar hydrogels prepared from bolaamphiphiles show a loss in structure (fiber to micelles) and elastic properties at temperatures rather in the order of 50-55°C.[53] These features make these materials more similar to silk fibroin, stable against temperature above 60°C,[65] or calcium-coordinated alginate,[66] than SAFIN hydrogels, and could be related to the stability of the metal-ion complex, as reported for silver stearate systems.[63]

If the above shows that both $Ca^{2+}$ and $Ag^+$ play an important structural role in the fiber's formation, with striking consequences on the elasticity and stability of the gels, complementary experiments show that their role is almost unique, if compared to other cations.[22] Alkaline earth, from $Mg^{2+}$ to $Ba^{2+}$, but also as transition metals, like $Fe^{2+}$, $Zn^{2+}$, $Mn^{2+}$ among others, can induce gelation[22] through a micelle-to-fiber transition.[48] However, the structure and morphology of the fibers, their size and aggregation state does not seem to be the same as the one found for $\{Ca^{2+}\}$G-C18:1 and $\{Ag^+\}$G-C18:1. In parallel, hydrogelation differs from cation to cation, it is either not induced or instantaneous but localized to the solution volume, where the ion solution is injected. We could find similar gelation with comparable structure of the fibers only with $Mn^{2+}$ or $Cr^{2+}$, the latter most likely oxidized into $Cr^{3+}$. Cations can differ in size, charge density, polarizability, polarizability, coordination sphere, reactivity, acido-basicity, just to cite some.[67–69] For this reason, a complete understanding of this new system would be challenging. Nonetheless, considering the pH-dependent speciation diagrams of each cation,[67] we found that hydrogels with lamellar raft domains are mostly observed for those cations, which exist as free ions in solution in the pH range of hydrogel formation and which can adopt non-octahedral coordination, thus reinforcing the importance of the non-isotropic coordination by G-C18:1.

Finally, the sugar headgroup could play an important role both in the packing of the lipid dimers and in the inter-fiber interactions. It was previously shown that G-C18:1 forms interdigitated membranes,[20,21] possibly constituted by flip-flopped molecules, so to maximize packing and minimize repulsive steric interactions among sugar headgroups. A similar phenomenon could occur inside the fibers and could naturally explain the negative charge density along their side. Conformation of the sugar headgroup in glycolipids has also been reported to have a strong impact on the structure of their corresponding supramolecular aggregation.[70,71] The difference in terms of coordination geometry between silver and calcium, associated to a conformational effect of glucose could eventually explain the differences in terms of packing of the G-C18:1 dimers and the overall stability of the fibers and corresponding hydrogels. To highlight again the unique behaviour of G-C18:1,[19] as other glycolipid amphiphiles with similar structure do not specifically react with mono and divalent cations.[23,25]

Finally, cation-mediated interactions between adjacent fibers could also occur upon adsorption of cations on the neutral glucose headgroups. This mechanism, although less likely occurring than electrostatic attraction or coordination chemistry, is not to be excluded, as shown by recent data on the adsorption of $Ca^{2+}$ and other ions onto neutral glycolipid membranes.[72] Nonetheless, we exclude this mechanism on the basis of $^{13}C$ ssNMR arguments, which do not show significant differences in the spectral region of glucose between micellar and fiber samples.[19]

**Conclusions**

SAFiN from low molecular weight compounds commonly form hydrogels under dilute conditions, generally in the order of 1 wt% or less. Whatever the stimulus (temperature, pH, ionic strength) that drives fibrillation and consequent hydrogel formation, the structure of the gel is generally very similar across samples of different origin. SAFiN hydrogels are formed by an entangled network of infinitely long fibers. The typical X-ray or neutron scattering profiles display the form, and sometimes structure, factor of individual fibers. In some specific cases, fibers form columnar structures (bundles) with a hexagonal order, driven by repulsive electrostatic interactions.

In this work, we employ a new bolaform glycolipid containing a free-standing COOH group and belonging to the broad family of biological amphiphiles (biosurfactants). In its micellar phase at pH above neutrality, addition of a $Ca^{2+}$ or $Ag^{+}$ solution drives a micelle-to-fiber phase transition, not expected for this specific compound. Fibers have an approximate cross-section of 10 and 5 nm, respectively for {$Ca^{2+}$}G-C18:1 and {$Ag^{+}$}G-C18:1 systems. Above about 0.5 wt%, addition of the ion solution also drives the formation of a hydrogel, of which the strength is maximized at a stoichiometric negative/positive charge ratio between the $COO^{-}$ group of the glucolipid and the cation. The gel strength (G´) increases with glucolipid concentration according to a power dependency in the order of 2.3-2.6. The strength of the hydrogels is one order of magnitude higher for {$Ag^{+}$}G-C18:1 with respect to {$Ca^{2+}$}G-C18:1 and so is the stability against shearing and temperature. The elastic properties of {$Ag^{+}$}G-C18:1 gels are kept as such up to at least 70°C, while the elasticity of $Ca^{2+}$ gels is lost below about 65°C.

Both {$Ca^{2+}$}G-C18:1 and {$Ag^{+}$}G-C18:1 gels have a unique structure, which most likely explains their elasticity and stability. {$Ag^{+}$}G-C18:1 hydrogels display a long-range lamellar order of the fibers, systematically observed at rest, after shear, upon heating and combining heating and shearing. Such order is observed both by cryo-TEM, showing how individual fibers spontaneously assemble into dimers or oligomers, and, above-all, SAXS and rheo-SAXS. The latter shows a distinct lamellar structure factor, of which the typical period corresponds to the inter-fiber distance. Similar results are also observed for $\{Ca^{2+}\}$G-C18:1 hydrogels, although the lamellar structure factor is partially or totally lost when hydrogels are sheared.

The structural SAXS and cryo-TEM data prone for a β-sheet-like raft structure of both $\{Ca^{2+}\}$G-C18:1 and $\{Ag^+\}$G-C18:1 hydrogels up to at least 5 wt% of glucolipid in solution. A similar structure, known as "nano-fishnet", is not known for SAFiN hydrogels but generally found for more complex aminoacid-based systems, like natural actin and silk proteins. The strong stability towards shear and temperature can probably be explained by such β-sheet-like structure, in analogy to silk fibroin, of which the hydrogels are extremely stable. This hypothesis is corroborated by the fact that the β-sheet-like structure is less stable for $\{Ca^{2+}\}$G-C18:1 hydrogels, which are softer by one order of magnitude and less stable towards temperature.

The main hypothesis explaining the existence of β-sheet-like rafts concerns the anisotropic distribution of negative charges in the fibers, most likely localized on the sides. Ions counterbalance the charge and generate a bridging effect between adjacent fibers. However, binding of calcium is reversible, thus generating metastable rafts, while silver binding is much more stable, generating long-lasting β-sheet-like rafts and, consequently, more stable hydrogels.

Compared to other glycolipid biosurfactants, G-C18:1 shows a unique self-assembly behaviour in the presence of $Ca^{2+}$ or $Ag^+$, as the phase behaviour of molecules with similar chemical structures is not affected at all by cations. Similarly, the "nano-fishnet" structure of G-C18:1 hydrogels is quite unique, even compared to most other amphiphiles found in the literature.

**Acknowledgements**

We thank Dr. S. Roelants at Gent University and Bio Base Europe Pilot Plant, Gent, Belgium for dealing with and shipping the G-C18:1 glycolipid. Authors kindly acknowledge the French ANR, Project N° SELFAMPHI - 19-CE43-0012-01. Soleil synchrotron is acknowledged for financial support during the beamtime associated to the proposal number N°20200532. ESRF synchrotron is acknowledged for financial support during the beamtime associated to the proposal numbers N° SC-4976 and MX 2311.

**Supporting Information**

Table S 1 present a literature survey on the SAFiN structure based on the scattering (SAXS or SANS) data reported in each cited article. Figure S 1 presents the elastic and viscous moduli {$Ca^{2+}$}G-C18:1 or {$Ag^+$}G-C18:1 hydrogels. Figure S 2 shows the time-dependency of the gelation of {$Ca^{2+}$}G-C18:1 hydrogels. Figure S 3 shows the SAXS spectra of a {$Ca^{2+}$}G-C18:1 hydrogels at a molar ratio 0.61 for different surfactant concentrations, at a) pH8, b) pH10. Figure S 4 shows the SAXS profiles recorded during temperature increase from 25°C to 70°C for a {$Ca^{2+}$}G-C18:1 gel. Figure S 5 shows the 2D SAXS spectra of {$Ca^{2+}$} or {$Ag^+$} G-C18:1 hydrogels at pH 10 recorded with a rheo-SAXS experiment.

**Figure S 1 - Elastic and viscous moduli, G' and G'', of a,b) {$Ca^{2+}$}G-C18:1 (red square) or c,d) {$Ag^+$}G-C18:1 (green triangle) hydrogels ($C_{G-C18:1}$= 3 wt%, pH 10). Ion concentrations in the lipid solution are 50 mM and 200 mM.**

**Figure S 2 – Time-dependency of the gelation ($\gamma$= 0.3%, f= 1 Hz) of {$Ca^{2+}$}G-C18:1 hydrogels (C= 3 wt%, pH 10). The charge ratio is [$Ca^{2+}$]/[G-C18:1]= 0.61. Hydrogel is left in a closed sample-holder and sampled at regular intervals, and not left under air on the rheometer geometry.**

**Figure S 3 – SAXS spectra of a {$Ca^{2+}$}G-C18:1 hydrogels at a molar ratio 0.61 for different surfactant concentrations, at a) pH8, b) pH10. The charge ratio is [$Ca^{2+}$]/[G-C18:1]= 0.61.**

**Figure S 4 - SAXS profiles recorded during temperature increase from 25°C to 70°C for a {$Ca^{2+}$}G-C18:1 gel at 3 wt%, basic pH and [$Ca^{2+}$]/[G-C18:1]= 0.61. Please note that the experimental setup in the SAXS-coupled rheometer (here, ID02, ESRF) allows a poor control over temperature.**

**Figure S 5 – Rheo-SAXS experiment. 2D SAXS spectra of {$Ca^{2+}$} (molar ratio 0.61) or {$Ag^+$} (molar ratio 1.0) G-C18:1 hydrogel ($C_{G-C18:1}$ = 3%) at pH 10 sheared at 0 s$^{-1}$ or 100 s$^{-1}$**

# Supporting information

# Shear recovery and temperature stability of Ca$^{2+}$ and Ag$^+$ glycolipid fibrillar metallogels with unusual β-sheet-like domains

Alexandre Poirier,[a] Patrick Le Griel,[a] Thomas Bizien,[b] Thomas Zinn,[c] Petra Pernot,[c] Niki Baccile[a,*]

[a] Sorbonne Université, Centre National de la Recherche Scientifique, Laboratoire de Chimie de la Matière Condensée de Paris, LCMCP, F-75005 Paris, France
[b] Synchrotron Soleil, L'Orme des Merisiers, Saint-Aubin, BP48, 91192 Gif-sur-Yvette Cedex, France
[c] ESRF – The European Synchrotron, CS40220, 38043 Grenoble, France

* Corresponding author:
Dr. Niki Baccile
E-mail address: niki.baccile@sorbonne-universite.fr
    Phone: +33 1 44 27 56 77

**Table S 1 – Literature survey based on the scattering (SAXS or SANS) data reported in each cited article. Articles cited under *SAFiN with disordered fibers* report a typical scattering profile of the fiber alone, with or without a structure peak. Articles cited under *SAFiN with suprafibrillar assembly* report scattering profiles where the fiber's form factor is superimposed with the structure factor associated to the 3D organization of the fibers. Articles cited under *β-sheet-like gel ("nano-fishnet")* report those hydrogels characterized by entanglement and β-sheet or β-sheet-like structure.**

| SAFiN with disordered fibers | |
|---|---|
| Not gelled | [1–4] |
| Gels | [5,6,15–17,7–14] |
| **SAFiN with suprafibrillar assembly** | |
| Hexagonal bundles/Columnar hexagonal | [18–23] |
| Raft-like/lamellar (solution, not gelled) | [24,25] |
| **β-sheet-like gel ("nano-fishnet")** | |
| Biopolymers (fibroin, actin) | [26–33] |
| SAFiN | This work |

The typical SAXS profiles of self-assembled fibrillar hydrogels from amphiphiles generally show an intense low-q scattering, of which the slope is related to the morphology (-1 for thin fibers and -2 for belts and ribbons[4]) and either a broad oscillation (form factor)[3] or a broad structural peak.[34] We are not aware of SAXS fingerprints of fibrillar hydrogels from low-molecular weight amphiphiles with a series of diffraction peaks in a lamellar order.

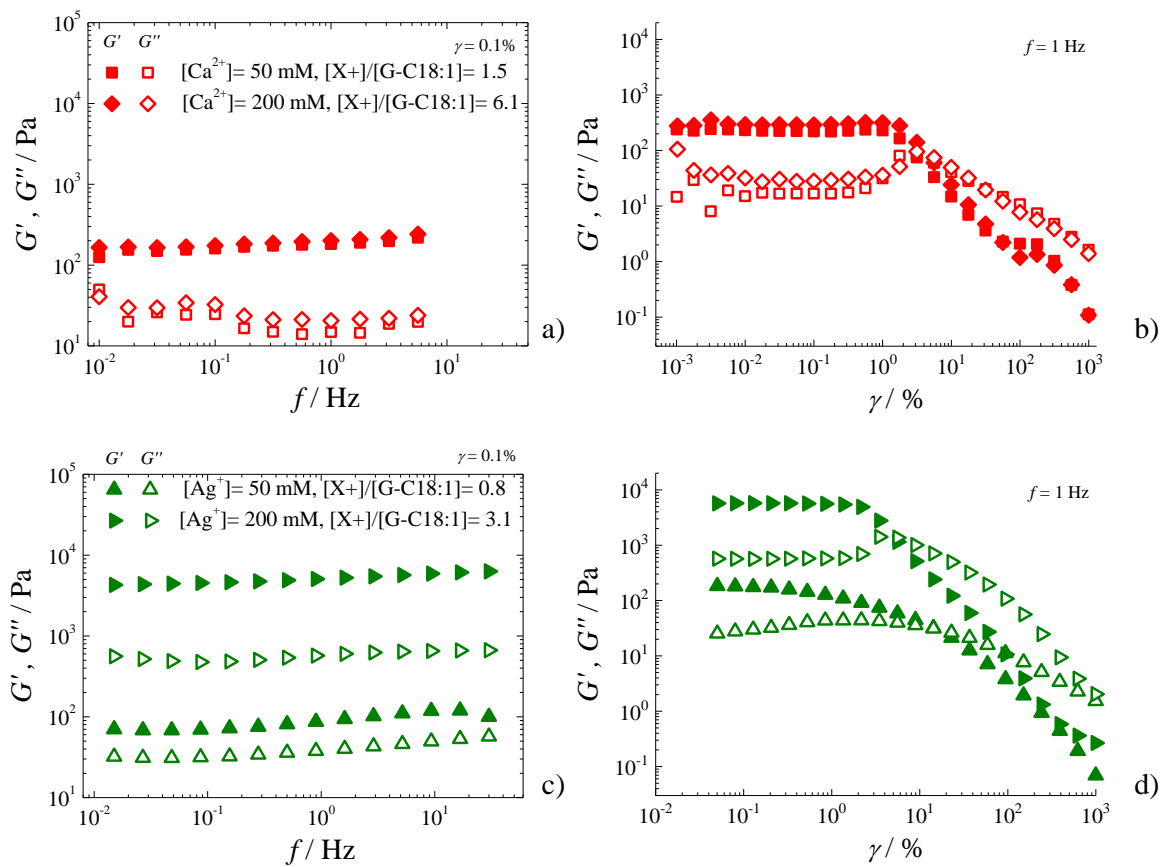

**Figure S 1 - Elastic and viscous moduli, G´ and G´´, of a,b) {Ca²⁺}G-C18:1 (red square and diamonds) or c,d) {Ag⁺}G-C18:1 (green triangle) hydrogels (C$_{G-C18:1}$= 3 wt%, pH 10). Ion concentrations in the lipid solution are 50 mM and 200 mM.**

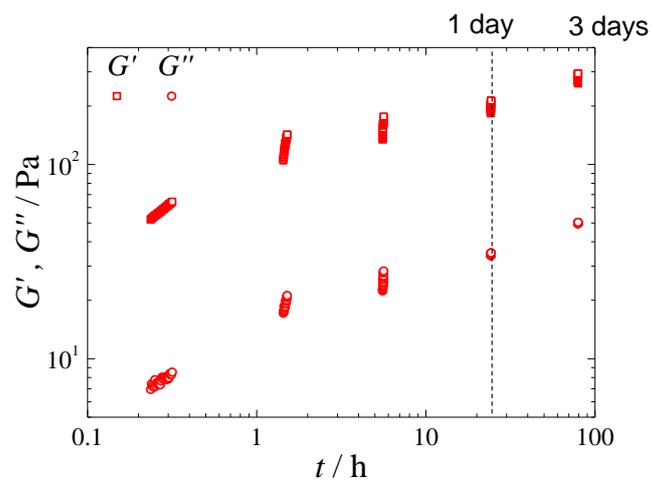

**Figure S 2 – Time-dependency of the gelation (γ= 0.3%, f= 1 Hz) of {Ca²⁺}G-C18:1 hydrogels (C= 3 wt%, pH 10). The charge ratio is [Ca²⁺]/[G-C18:1]= 0.61. Hydrogel is left in a closed sample-holder and sampled at regular intervals, and not left under air on the rheometer geometry.**

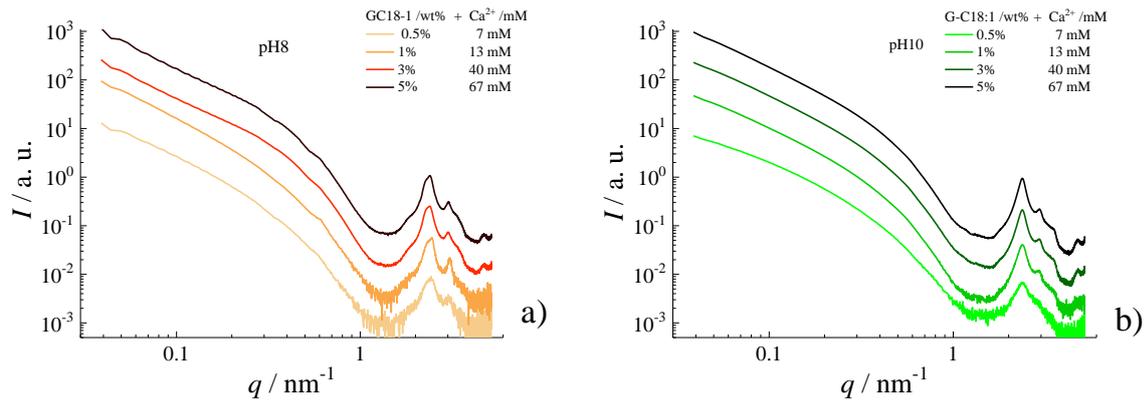

**Figure S 3 – SAXS spectra of a {Ca²⁺}G-C18:1 hydrogels at a molar ratio 0.61 for different surfactant concentrations, at a) pH8, b) pH10. The charge ratio is [Ca²⁺]/[G-C18:1]= 0.61. Data are shifted by a factor of about 5.**

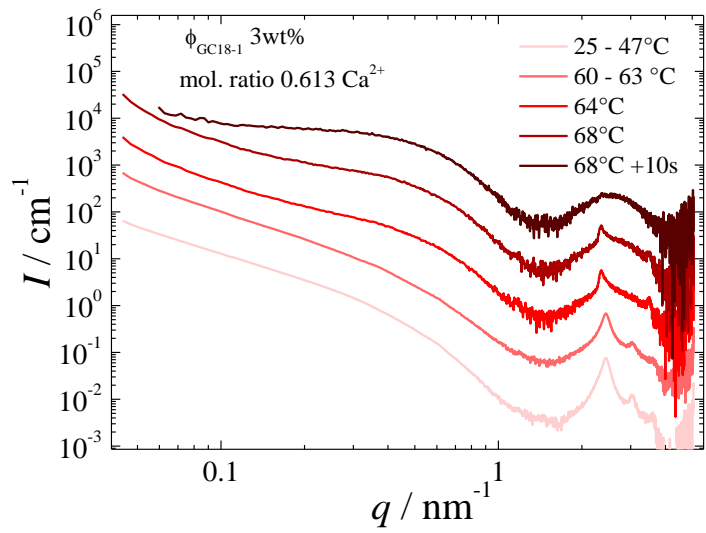

**Figure S 4 - SAXS profiles recorded during temperature increase from 25°C to 70°C for a {Ca²⁺}G-C18:1 gel at 3 wt%, basic pH and [Ca²⁺]/[G-C18:1]= 0.61. Please note that the experimental setup in the SAXS-coupled rheometer (here, ID02, ESRF) is not suitable for a precise control over temperature. Data are shifted by a factor of about 15.**

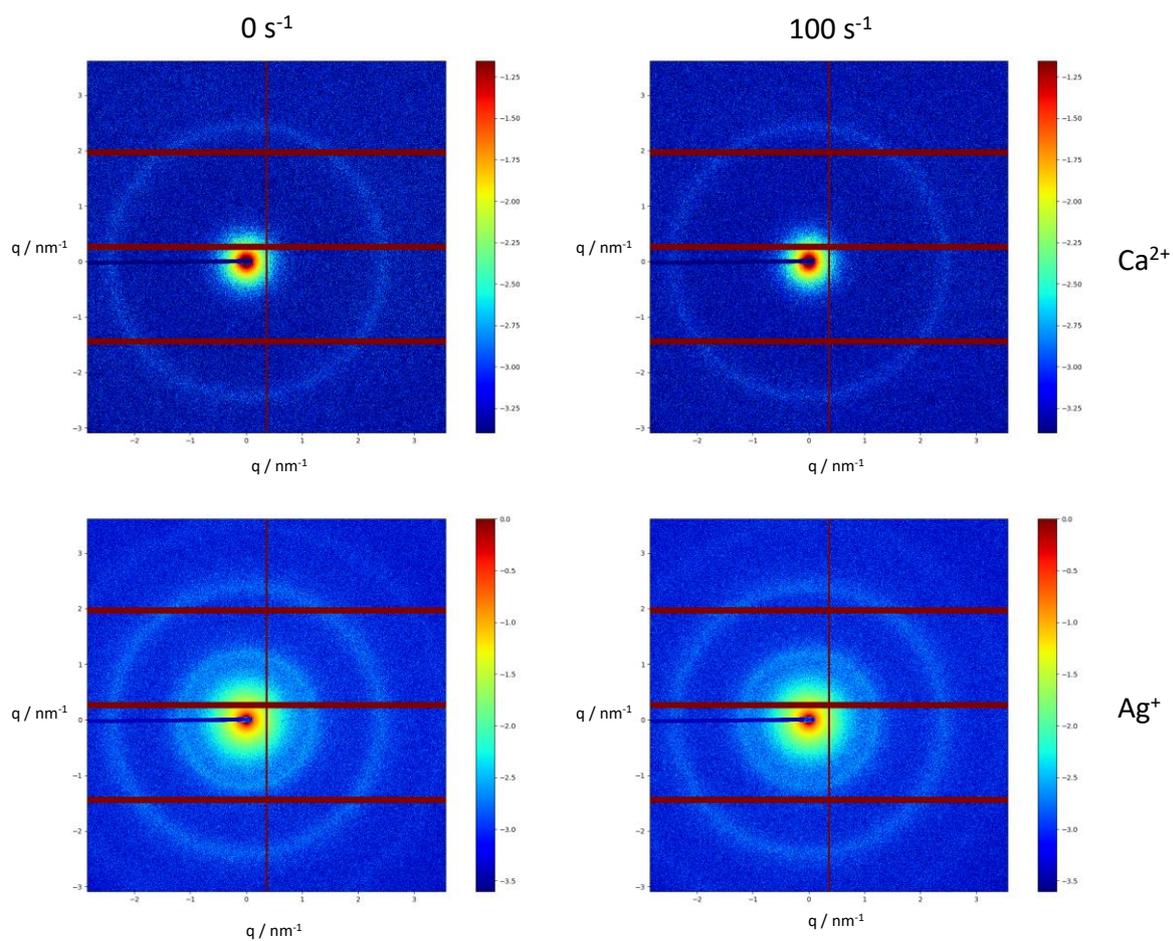

**Figure S 5 – Rheo-SAXS experiment. 2D SAXS spectra of {Ca²⁺} (molar ratio 0.61) or {Ag⁺} (molar ratio 1.0) G-C18:1 hydrogel ($C_{G-C18:1}$ = 3%) at pH 10 sheared at 0 s⁻¹ or 100 s⁻¹**